\documentclass[aps,prl,floatfix,twocolumn,superscriptaddress,longbibliography,reprint, showpacs]{revtex4-1}

\usepackage[%
	colorlinks=true,
	urlcolor=blue,
	linkcolor=blue,
	citecolor=blue,
	breaklinks=true
]{hyperref}
\usepackage[english]{babel} 
\usepackage{bm}
\usepackage{amsmath}
\usepackage{graphicx} 
\graphicspath{ {./Figures/} } 
\usepackage{color}
\usepackage{siunitx} 
\usepackage[all]{hypcap} 
\usepackage[normalem]{ulem} 

\def\degree{\ensuremath{^{\circ}}}

\begin{document}

\title{Large polarons as key quasiparticles in SrTiO$_{3}$ and SrTiO$_{3}$-based heterostructures}

\author{Andrey~Geondzhian}
\affiliation{ESRF -- The European Synchrotron, 71 Avenue des Martyrs, CS 40220, F-38043 Grenoble, France}

\author{Alessia~Sambri}
\affiliation{CNR-SPIN Complesso Monte-Santangelo via Cinthia, I-80126 Napoli, Italy}

\author{Gabriella~M.~De~Luca}
\affiliation{Dipartimento di Fisica ``Ettore Pancini'' Universit\`a di Napoli ``Federico II'', Complesso Monte-Santangelo via Cinthia, I-80126 Napoli, Italy}
\affiliation{CNR-SPIN Complesso Monte-Santangelo via Cinthia, I-80126 Napoli, Italy}

\author{Roberto~Di~Capua}
\affiliation{Dipartimento di Fisica ``Ettore Pancini'' Universit\`a di Napoli ``Federico II'', Complesso Monte-Santangelo via Cinthia, I-80126 Napoli, Italy}
\affiliation{CNR-SPIN Complesso Monte-Santangelo via Cinthia, I-80126 Napoli, Italy}

\author{Emiliano Di Gennaro}
\affiliation{Dipartimento di Fisica ``Ettore Pancini'' Universit\`a di Napoli ``Federico II'', Complesso Monte-Santangelo via Cinthia, I-80126 Napoli, Italy}
\affiliation{CNR-SPIN Complesso Monte-Santangelo via Cinthia, I-80126 Napoli, Italy}

\author{Davide Betto}
\altaffiliation[Present address: ]{Max Planck Institut f\"{u}r Festk\"{o}rperforschung, Heisenbergstrasse 1, D-70569 Stuttgart, Germany}
\affiliation{ESRF -- The European Synchrotron, 71 Avenue des Martyrs, CS 40220, F-38043 Grenoble, France}

\author{Matteo Rossi}
\affiliation{Dipartimento di Fisica, Politecnico di Milano, piazza Leonardo da Vinci 32, I-20133 Milano, Italy}
\affiliation{Stanford Institute for Materials and Energy Sciences, SLAC National Accelerator Laboratory, 2575 Sand Hill Road, Menlo Park, CA}

\author{Ying Ying Peng}
\affiliation{Dipartimento di Fisica, Politecnico di Milano, piazza Leonardo da Vinci 32, I-20133 Milano, Italy}
\affiliation{International Center for Quantum Materials,School of Physics, Peking University, Beijing 100871, China}

\author{Roberto Fumagalli}
\affiliation{Dipartimento di Fisica, Politecnico di Milano, piazza Leonardo da Vinci 32, I-20133 Milano, Italy}

\author{Nicholas~B.~Brookes}
\affiliation{ESRF -- The European Synchrotron, 71 Avenue des Martyrs, CS 40220, F-38043 Grenoble, France}

\author{Lucio~Braicovich}
\affiliation{ESRF -- The European Synchrotron, 71 Avenue des Martyrs, CS 40220, F-38043 Grenoble, France}
\affiliation{Dipartimento di Fisica, Politecnico di Milano, piazza Leonardo da Vinci 32, I-20133 Milano, Italy}

\author{Keith~Gilmore}
\affiliation{ESRF -- The European Synchrotron, 71 Avenue des Martyrs, CS 40220, F-38043 Grenoble, France}
\affiliation{Condensed Matter Physics and Materials Science Division, Brookhaven National Laboratory, Upton, NY 11973-5000, USA}

\author{Giacomo~Ghiringhelli}
\affiliation{Dipartimento di Fisica, Politecnico di Milano, piazza Leonardo da Vinci 32, I-20133 Milano, Italy}
\affiliation{CNR-SPIN, Piazza Leonardo da Vinci 32, I-20133 Milano, Italy}

\author{Marco Salluzzo}
\email{marco.salluzzo@spin.cnr.it}
\affiliation{CNR-SPIN Complesso Monte-Santangelo via Cinthia, I-80126 Napoli, Italy}

\date{\today}

\begin{abstract}
Despite its simple structure and low degree of electronic correlation, SrTiO$_3$ (STO) features collective phenomena linked to charge transport and, ultimately, superconductivity, that are not yet fully explained. Thus, a better insight in the nature of the quasiparticles shaping the electronic and conduction properties of STO is needed. We studied the low energy excitations of bulk STO and of the LaAlO$_{3}$/SrTiO$_{3}$ two dimensional electron gas (2DEG) by Ti L$_3$ edge resonant inelastic x-ray scattering. In all samples, we find the hallmark of polarons in the form of intense $dd$+phonon excitations, and a decrease of the LO3-mode electron-phonon coupling when going from insulating to highly conducting STO single crystals and heterostructures. Both results are attributed to the dynamic screening of the large polaron self-induced polarization, showing that the low temperature physics of STO and STO-based 2DEGs is dominated by large polaron quasiparticles.
\end{abstract}


\maketitle

\begin{figure}
\includegraphics[width=0.48\textwidth]{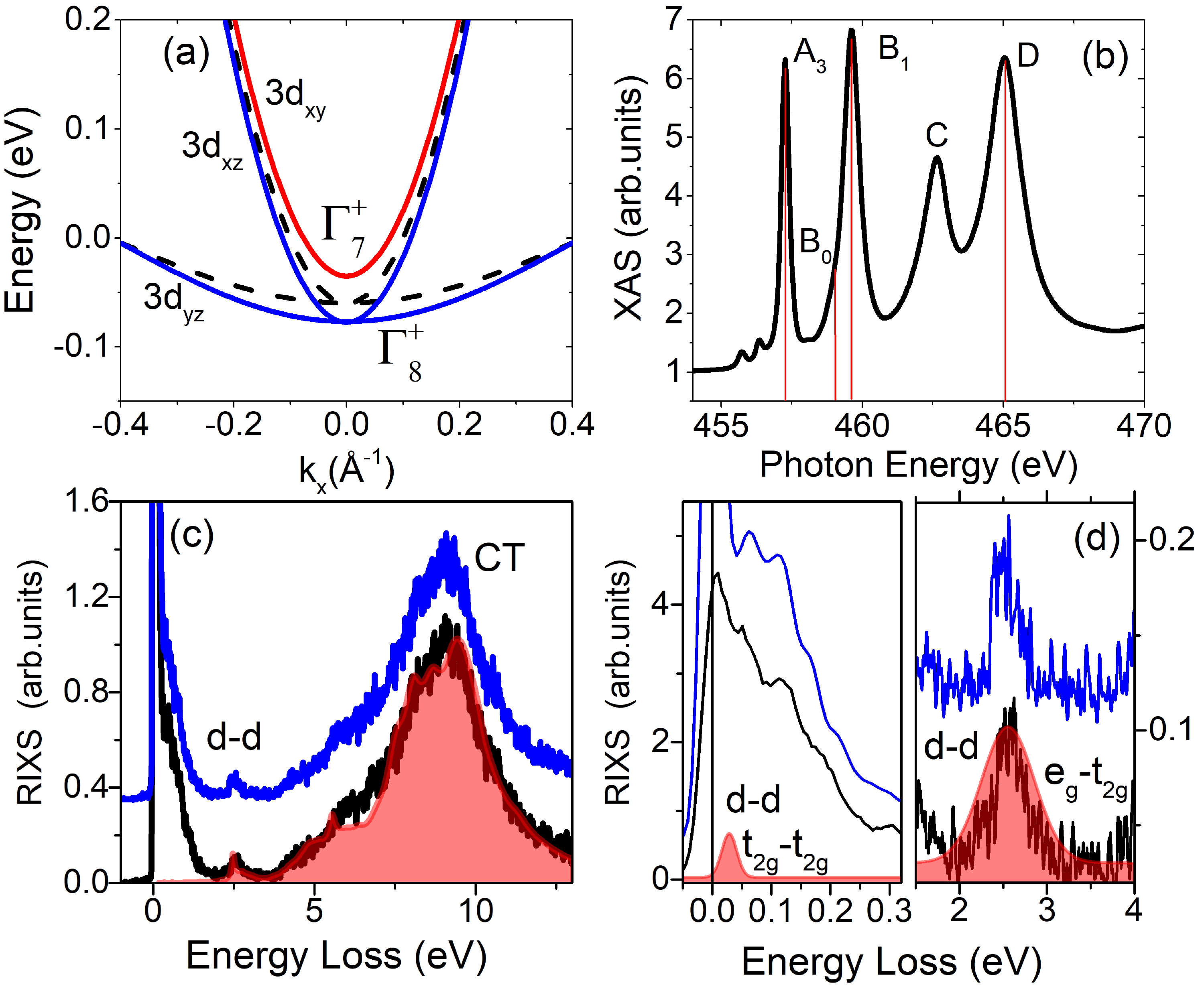}
\caption{\label{Fig:overview}Overview of XAS and RIXS spectra: (a) Electronic structure of bulk STO derived from tight binding calculations~\cite{Zhong2013} without (dashed black line) and with SOC (blue lines, $\Gamma_8^+$ quartet, red $\Gamma_7^+$ doublet). (b) XAS spectra on conducting bulk STO. (c) RIXS spectra at B1, normalized to the maximum of the charge transfer (CT) peak, for insulating STO (black line) and LAO/STO bilayer (blue line). The filled red shaded region is the calculated RIXS spectra using a BSE approach reproducing both $dd$ (peak around 2.5~eV) and CT excitations. (d) Expanded view of the low-(left panel) and mid-energy (right panel) regions, and atomic multiplet calculations including spin orbit interaction, which show the additional intra-t$_{2g}$ $dd$ peak around 30~meV.}
\end{figure}

After the discovery of a high mobility and superconducting two-dimensional electron gas (2DEG) at the LaAlO$_{3}$/SrTiO$_3$ (LAO/STO) interface \cite{Ohtomo2004,Reyren2007}, and of high T$_c$ superconductivity in CaCuO$_2$/SrTiO$_3$ \cite{DiCastro2012} and FeSe/SrTiO$_3$  \cite{Wang2012, Ge2015, Lee2014} bilayers, work on bulk and surface electronic properties of SrTiO$_{3}$ (STO) received renewed interest. Despite a simple band structure, the normal and superconducting properties of STO and STO-based heterostructures are not yet fully understood. In the bulk, the $3d$-Ti $t_{2g}$ ($3d_{xy}$, $3d_{xz}$, $3d_{yz}$) manifold (Fig.~\ref{Fig:overview}a) forms three uppermost bands characterized by heavy and light effective masses along the primitive lattice vectors \cite{Mattheiss1972, Zhong2013}. Spin-orbit coupling (SOC) removes the degeneracy, and the six-fold $t_{2g}$ bands are split by about 30~meV in a $\Gamma_7^+$ doublet and a $\Gamma_8^+$ quartet \cite{Zhong2013}. These new bands, formed by the mixing of atomic $3d$-Ti $t_{2g}$ states, still partially retain their overall orbital character (Fig.~\ref{Fig:overview}a). The band hierarchy is reversed, due to confinement, in the 2DEG at the (001) LAO/STO interface \cite{Salluzzo2009,Cancellieri2014}, where bands with prevalent $3d_{xy}$ orbital character are lower in energy.

Optical spectroscopy \cite{vanMechelen2008}, transport studies \cite{vanderMarel2011} and angle resolved photoemission(ARPES) \cite{Meevasana2010} on Nb-doped STO showed that at moderate electron density, $n_{3D}\leq10^{20}$~ cm$^{-3}$, the carrier effective mass is larger than the free electron value and decreases with doping. These results point towards a key role played by the coupling of electrons to optical phonon modes, thus forming large polarons \cite{Devreese2010}. Although most of the features observed in the optical conductivity spectra can be explained by a large polaron model \cite{Devreese2010, Klimin2020}, there is still no consensus about the type of quasiparticle effectively determining the electronic properties of doped STO and the mechanism of superconductivity. Several models were proposed, including pairing mediated by optical phonons \cite{Baratoff1981,Appel1969}, condensation of large polarons \cite{Klimin2014}, or other exotic pairings in a Fermi liquid, such as the recent proposal of superconductivity mediated by ferroelectric fluctuations \cite{Edge2015}.

Here, we use high resolution Ti L$_3$ edge resonant inelastic x-ray scattering (RIXS) to probe elementary low energy excitations in insulating and conducting STO and LAO/STO heterostructures. We find that the electron-phonon coupling (EPC) to the longitudinal $\sim $100~meV optical phonon mode, LO3, decreases as a function of the carrier density for both $t_{2g}$ and $e_{g}$ electrons. More importantly, we observe a $\sim$130~meV composite excitation assigned to an intra $t_{2g}$ $dd$ transition accompanied by the emission of an LO3 phonon, providing evidence of large polaron quasi-particles in bulk STO and in the LAO/STO 2DEG.

RIXS experiments were performed at beamline ID32 of the European Synchrotron Radiation Facility. Thanks to the new RIXS setup \cite{Brookes:2018dz}, measurements could be performed with a resolution three times better than previous studies on other titanates and on the LAO/STO system \cite{Moser2015,Fatale2016,Zhou2011,Pfaff2018}. We studied four types of samples, characterized by different carrier density, namely insulating and conducting STO single crystals, LAO/STO bilayers and a LAO/STO multilayer composed by eight repetitions of a LAO(10 unit cells)/STO(10 unit cells) bilayer grown by pulsed laser deposition (Table~\ref{tab:samples}) \cite{Suppl}. The RIXS spectra were acquired at 20K with incoming photons at selected energies across the Ti L$_3$ edge (Fig.~\ref{Fig:overview}b) labelled as A3, B0 (1.6~eV above A3) and B1. In a simplified scenario for a Ti$^{4+}$ ion in $3d^0$ configuration, the A3 and B1 absorption peaks correspond to the excitation of a $2p_{3/2}$ core electron into a $3d$ state of $t_{2g}$ and $e_g$ symmetry, respectively. The B0 energy, apparently at the bottom of a valley in the absorption spectrum of STO, corresponds to a peak for a Ti$^{3+}$ ion in $3d^{1}$ configuration. The incident photon polarization was set perpendicular to the scattering plane ($\sigma$-pol) and at a scattering angle of 149.5\degree, which corresponds to the point (-0.2,0,0.2) in the reciprocal lattice. The resolution was between 30 and 35~meV (full width at half maximum), estimated from the fit of the elastic peak measured on silver contacts on the samples.

\begin{table}[b]
\caption{\label{tab:samples}Summary of the sample types studied in this work and their 10 K carrier density and resistivity.}
\begin{ruledtabular}
\begin{tabular}{lrl}
\textrm{Sample}&
\textrm{n$_{3D}$(cm$^{-3}$)}&
\textrm{$\rho$(m$\Omega$cm)}\\
\colrule
STO insulating & \textless 10$^{15}$ & \textgreater 10$^{5}$\\
STO conducting & 5x10$^{19}$ & 0.2\\
LAO/STO bilayer & 2-3x10$^{19}$ & 0.4-0.5\\
LAO/STO multilayer & 0.5-1x10$^{21}$ & 0.01\\
\end{tabular}
\end{ruledtabular}
\end{table}

A typical RIXS spectrum at B1 consists of a broad charge transfer (CT) band between 4~eV and 14~eV, a very narrow peak around 2.5~eV due to inter-band $dd$ transitions, and low energy excitations, such as phonons, near the elastic peak (Fig.~\ref{Fig:overview}c-d). The $dd$ feature around 2.5~eV is observed in all the samples, even in the (nominally) undoped and insulating STO bulk single crystal, suggesting the presence of some $3d^{1}$ (Ti$^{3+}$) electrons \cite{Note}. The shape and relative intensities of the $dd$ excitation and the CT peaks for the insulating STO spectra are very well reproduced by RIXS cross section calculations, based on the Bethe-Salpeter equation (BSE)  \cite{Gilmore:2015fp},  that consider both $3d^{0}$ and $3d^{1}$ contributions in a 20 to 1 ratio. In order to evaluate the role of SOC on the multiplet spectra, we  used atomic multiplet calculations assuming the same fraction of $3d^{1}$ (Fig.~\ref{Fig:overview}d), which predicts an additional intra-$t_{2g}$ $dd$ peak around 30~meV energy loss, but no other feature up to 2.5~eV (see ref.~\cite{Suppl} ).

\begin{figure}[t]
\includegraphics[width=0.47\textwidth]{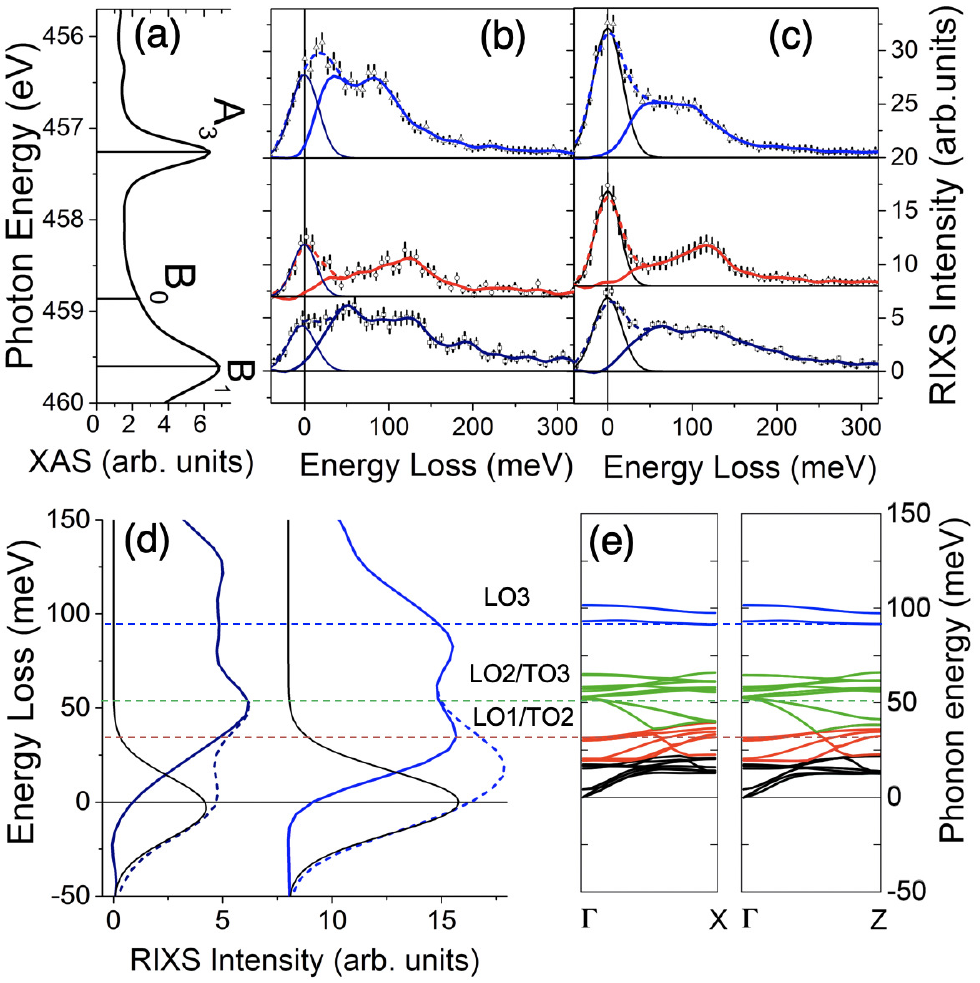}
\caption{\label{Fig:RIXS} Low energy excitation RIXS spectra: (a) XAS spectrum around Ti-L$_{3}$ edge of STO. (b)-(c) RIXS data as function of incoming photon energy along the XAS Ti-L$_{3}$ absorption edge for bulk (b) insulating and (c) conducting STO. For each energy, we show the raw scatter data, the elastic peak fit (black lines), and a 2 points fast Fourier transform (FFT) smoothing of the raw (short dashed lines) and elastic contribution subtracted data (solid lines).  Different colors correspond to different photon energies, namely A3 (blue), B0 (red) and B1 (dark blue) (from top to bottom). The elastic peak has been determined by fixing the instrumental resolution from reference spectra on silver. (d) 2 points FFT smoothed raw (dashed lines) and elastic subtracted (solid lines) data at A3 (blue) and B1 (dark blue) for the STO insulating sample. Dashed horizontal lines indicate the three main phonon peaks compared to (e) tabulated STO phonon dispersions \cite{Petretto2018,*Miranda2014}.}
\end{figure}
Figure \ref{Fig:RIXS} shows typical low energy loss RIXS data on insulating and conducting bulk STO for excitations at the A3, B0 and B1 energies. Three main features can be identified in the A3 and B1 spectra (Fig.~\ref{Fig:RIXS}d): at low, intermediate and high energies of 25-30~meV ($\omega_1$), 55-65~meV ($\omega_2$) and 90-100~meV ($\omega_3$), respectively. As shown in Fig.~\ref{Fig:RIXS}e, these energies match the known phonon branches of longitudinal (transverse) optical modes, namely LO1 (TO2), LO2 (TO3) and LO3. Besides the three main phonon peaks, several additional higher energy features are visible in the data, some of them occurring at multiples of $\omega_3$ ($\sim$200 and 300~meV) which correspond to two- and three-LO3-phonon replicas. Moreover, in the spectra at B1 and at B0, an additional particularly strong peak is visible around 125-135~meV in both insulating and conducting samples.

The data show strong differences in the low energy spectra at A3 and B1, where phonons are expected to couple mainly to the $t_{2g}$ and $e_{g}$ electrons, respectively. In order to determine the $t_{2g}$ and $e_{g}$ coupling strengths ($g_{t_{2g}}$ and $g_{e_{g}}$, respectively), it is necessary to take into account that the A3 (B1) XAS peak has not a pure $2p^5 3dt_{2g} ^1$ ($2p^5 3de_g ^1$) character, but rather contains an approximate 25\% mixing of $e_{g}$ (t$_{2g}$) orbital character \cite{Ogasawara2001, Gilmore2010}. Due to the above points, we performed simultaneous constrained fits of the A3 and B1 RIXS spectra employing a three-mode generalization of the Franck-Condon model \cite{Ament2011}. The model, described in details in Refs.~\cite{Suppl,Geondzhian:2020tn}, considers intermediate and final states containing not only multiples of a single mode but also mixed-mode double excitations. The Ti $2p$ core hole lifetime, $\Gamma$, was fixed to 110~meV (HWHM) for both A3 and B1 spectra \cite{Krause1979} (see Ref.~\cite{Suppl} for an analysis of the Ti $2p$ core-hole lifetime).

\begin{figure} [t]
\includegraphics[width=0.38\textwidth]{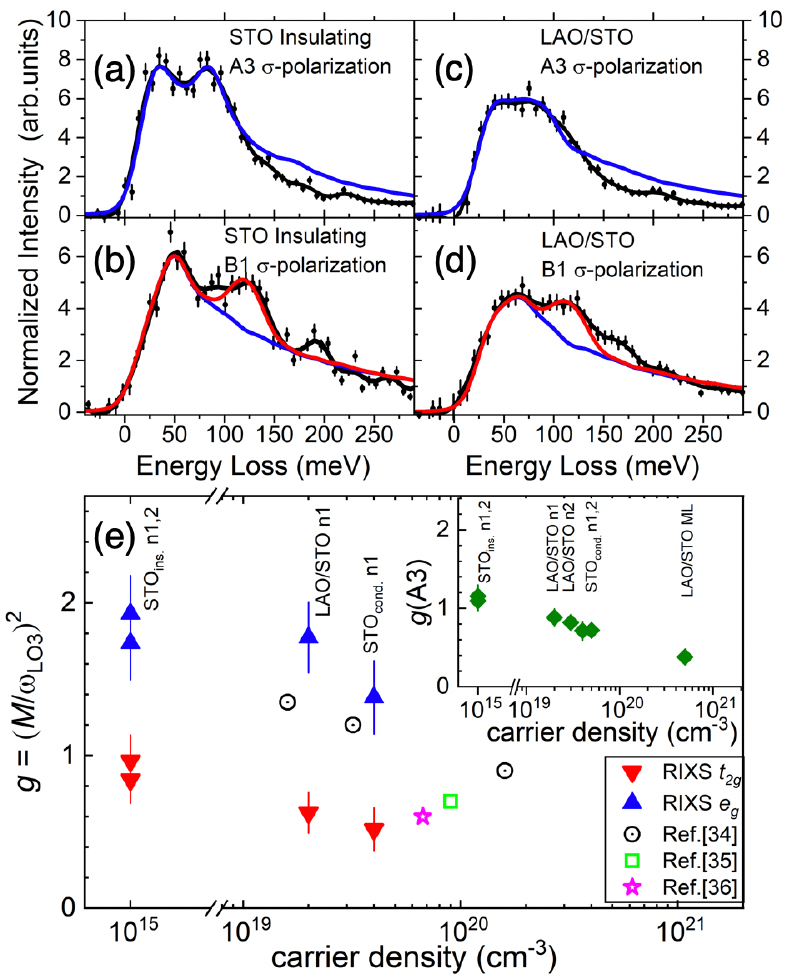}
\caption{\label{Fig:FIT} Fit of the RIXS spectra: Comparison between RIXS data (black circles) on insulating STO  ((a) and (b)) and conducting LAO/STO ((c) and (d)) and their fit at A3 (upper panels) and B1 (bottom panels). Black lines are 2 points FFT smoothing of the data. Blue lines are the fit using the phonon mixing model. Red lines are the fit including the intra-$t_{2g}$ $dd$ plus LO3 phonon RIXS cross section of Eq.~\ref{eq:RIXS}. (e) LO3 $g(t_{2g}$) (red triangles) and $g(e_{g}$) (blue triangles) EPC estimated from the analysis of the RIXS data at A3 and B1 as function of the carrier density $n_{3D}$. The results are compared to data from literature~\cite{Swartz2018,Wang2016,Cancellieri2016}. We assigned a doping of 10$^{15}$ cm$^{-3}$ to insulating STO, while $n_{3D}$ for STO in Ref.~\cite{Wang2016} and LAO/STO in Ref.~\cite{Cancellieri2016} is estimated from the reported 2D carrier density assuming a thickness of 10 nm of the 2DEG. The inset shows $g(A3)$ as function of $n_{3D}$ for all the samples studied. n1 and n2 designate different samples of the same type.}
\end{figure}

In Fig.~\ref{Fig:FIT}a-d we show the fits at A3 and B1 for insulating bulk STO and an LAO/STO bilayer (other data fit on different samples and a complete list of the fitting parameters are shown in supplementary materials.~\cite{Suppl}). We underline that most of the spectral weight above 100~meV and the very long tail in the data (extending above 0.5~eV) cannot be reproduced without considering phonon mixed excitations \cite{Suppl}, indicating the importance of using a mixed phonon model for a better fitting when two or more phonons have comparable electron phonon coupling and energies. We find $g_{e_{g}} > g_{t_{2g}}$ as a consequence of the larger spatial overlap of $e_{g}$ orbitals with neighboring oxygen $2p$ states leading to $\sigma$ bonds with respect to the $\pi$-bond forming $t_{2g}$ states. In Fig.~\ref{Fig:FIT}e, we show  the volume carrier density dependence of the LO3 electron phonon coupling obtained from our RIXS data, compared to other experimental data collected from literature \cite{Swartz2018,Wang2016,Cancellieri2016}. The analysis shows that the LO3 $g_{t_{2g}}$ EPC found by RIXS is in reasonable quantitative agreement with ARPES data reported on conducting STO  \cite{Wang2016}and LAO/STO \cite{Cancellieri2016} samples with similar doping, and that it decreases as function of the carrier density, a result consistent with the dynamical screening of large polaron quasiparticles self polarization. This trend is confirmed in the carrier density dependence of the LO3 EPC determined from the fit of the RIXS spectra at A3 on all the STO and LAO/STO samples investigated, shown in the inset of Fig.~\ref{Fig:FIT}e. In particular, the LAO/STO ML, which is characterized by the highest carrier density among the studied samples, exhibits a value of g(A3) of 0.4, consistent with the bare undressed EPC of doped STO.

Although being in agreement with the general features, the fitting cannot account for the extra peak at $\sim$130~meV in the B0 and B1 RIXS spectra, which in some cases is stronger than the single phonon features. Thus, it cannot be reproduced by the phonon mixing model and cannot be assigned to a multiple phonon replica. A pure $dd$ excitation is also unlikely, as it would correspond to a very large splitting of the $t_{2g}$ states even in bulk STO. Moreover, its energy position is consistent with a composite excitation including an intra-t$_{2g}$ transition for electrons in $3d^{1}$ Ti orbitals and one high energy (90-100~meV) LO3 optical phonon. This is expected in a system where electrons get dressed by the polar distortions of the lattice, thus forming polarons.

In order to verify this idea, we included in the fit the RIXS cross section from a composite $dd$ plus LO3 ($\omega_3$) phonon excitation given, following Ref.~\cite{Ament2010}, by:
\begin{equation}\label{eq:RIXS}
 \frac{d^2\sigma}{dEd\Omega}=\frac{|T_{dd}|^2}{\Gamma^2}e^{-g_{dd}} \sum_{n=1}^{\infty} \frac{g_{dd}^{\quad n}}{n!}\delta(E-E_{dd}-n\hbar\omega_3)
\end{equation}
where $T_{dd}$ is the polarization factor corresponding to the specific $dd$ excitation and $g_{dd}$ = $\left(M_{dd}/\omega_3\right)^2$ is the coupling constant, with $M_{dd}$ half of the Jahn-Teller energy $E_{JT}$. As shown in Fig.~\ref{Fig:FIT}b-d, a very satisfactory fitting of the data is obtained by only adjusting the value of $g_{dd}$ and a scaling factor (inclusive of the term $|T_{dd}|^2$ in Eq.~\ref{eq:RIXS}), without changing the phonon-related part.

\begin{figure}
\includegraphics[width=0.38\textwidth]{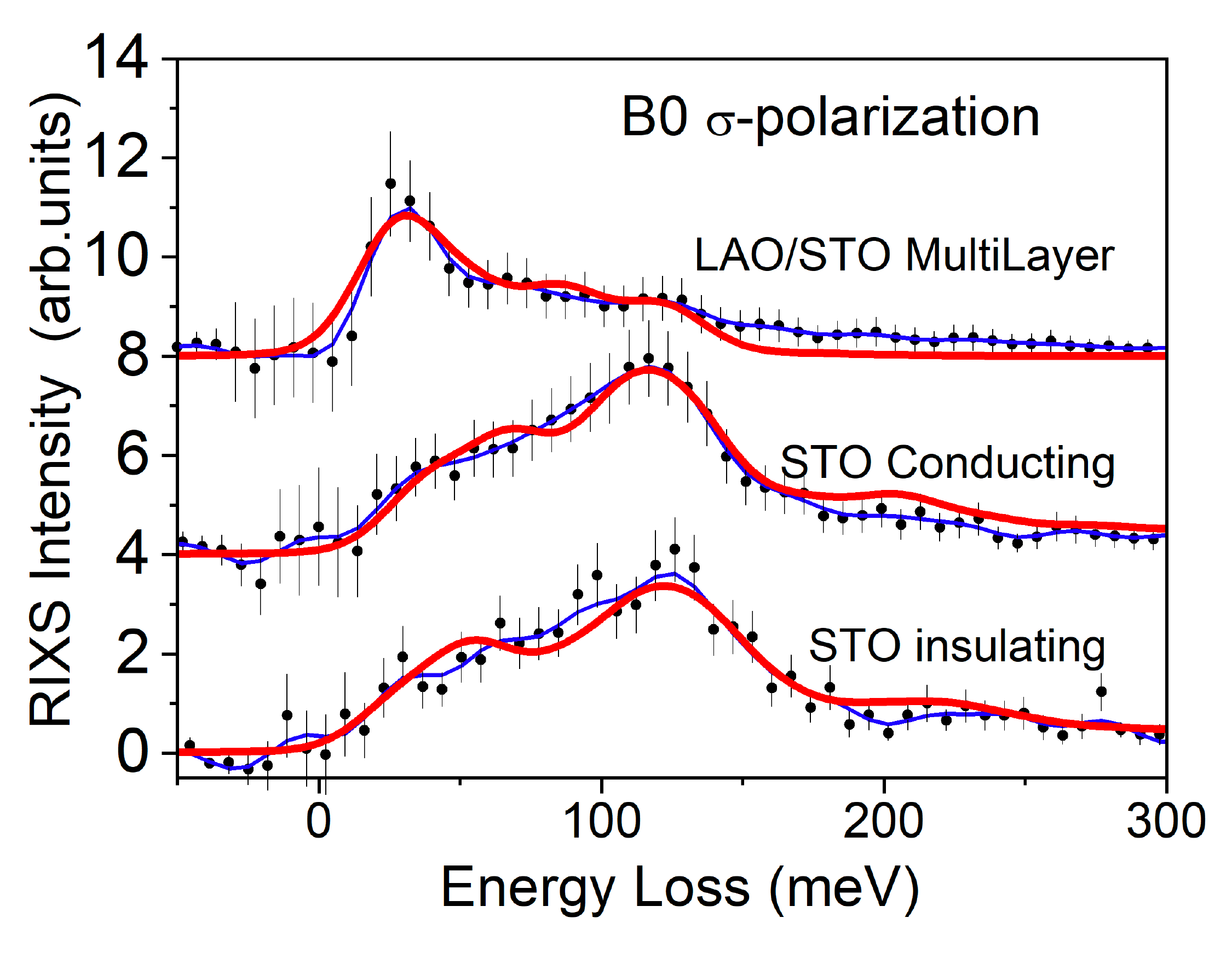}
\caption{\label{Fig:B0fit}Fit of the RIXS data at B0 for STO insulating, STO conducting and highly doped LAO/STO multilayer (data vertically displaced for clarity).}
\end{figure}

A further confirmation of this analysis comes from a fit of the B0 RIXS spectra. Here, the $\sim$130~meV peak is enhanced with respect to pure phonon excitations, because B0 is a resonant transition for $3d^{1}$ states. We are able to fit the B0 RIXS spectra on each sample using the same fitting parameters obtained at B1 by only adjusting a relative scaling factor for the purely phonon- and composite-, $dd$ plus LO3 phonon (Eq.~\ref{eq:RIXS}), excitation terms (Fig.~\ref{Fig:B0fit}). The value of $g_{dd}$ obtained from the fit at B0 and at B1 is 0.35 for insulating STO and 0.32 for conducting STO, and corresponds to $E_{JT} \simeq 105$~meV, a value of the same order of magnitude of $E_{JT}$ theoretically estimated for STO~\cite{Gilmore2010} and for BaTiO$_3$~\cite{Bersuker2015}. The same analysis on the highly doped LAO/STO multilayer, characterized by a coupling to the LO3 mode of the order of the bare, undressed, EPC, shows that at high carrier density the $\sim$130~meV feature is much weaker(Fig.~\ref{Fig:B0fit}). Indeed, from the fit at B0 we get  a value of  $g_{dd} \simeq 0.03$, reinforcing the interpretation of the $\sim$130~meV peak as an hallmark of large polaron quasiparticles and their disappearance at large carrier density.

Our results carry several outcomes. High resolution RIXS spectra not only quantify trends in the EPC for optical phonons, they also reveal polaronic excitations in bulk STO and STO based heterostructures, demonstrating that RIXS can be used as a new technique to investigate polaron physics both in insulating and conducting materials. We note that a $\sim$130~meV feature was observed in optical conductivity data of doped STO \cite{Devreese2010}, and recently explained within a large-polaron model by the inclusion of the dynamical screening of electrons from the lattice polarization \cite{Klimin2020}. The coincidence in energy position and carrier density dependence, with a disappearance at high doping, suggests that RIXS and optical conductivity observations have a common origin. Furthermore, by RIXS we have  shown that the $\sim$130~meV peak involves $dd$ intra-$t_{2g}$ transitions of $3d^{1}$ Ti$^{3+}$ states, accompanied by the excitation of LO3 phonons. Consequently, $t_{2g}$ electrons in STO form large polaron quasi-particles. Beside confirming earlier signatures provided by ARPES on the surface of STO \cite{Wang2016}, on LAO/STO \cite{Cancellieri2016} and on FeSe/STO bilayers \cite{Ge2015}, our study demonstrates more generally the emergence of large polaron physics in both bi-dimensional and three-dimensional titanates.

Finally, it emerges that polarons are observed also in nominally undoped STO, with a coupling constant well below the value expected for small polarons formation. Consequently, we can infer that even at the very low doping level, as that induced by residual defects or by long living photodoped  carriers, 3d${^1}$ electrons are dressed by long-range polar distortions of the lattice, as theoretically predicted in other wide band-gap materials like LiF \cite{Sio:2019jw}. Future investigations and theoretical modelling of the normal and superconductong state of STO and STO-based heterostructures will have to take in consideration the central role of large polarons in these materials.
\begin{acknowledgments}
The authors gratefully acknowledge the help of the ID32 beam line staff at ESRF. This project received funding from the EU Horizon H2020 project QUANTOX (grant 731473); from MIUR of Italy for the PRIN projects TOP-SPIN (Grant No. PRIN 20177SL7HC) and QUANTUM 2D (Grant No. PRIN 2017Z8TS5B); from the Fondazione CARIPLO and Regione Lombardia for the ERC-P-ReXS project (Grant No.~2016-0790). KG was supported by the U.S. Department of Energy, Office of Science, Basic Energy Sciences as part of the Computational Materials Science Program.
\end{acknowledgments}


\begin{thebibliography}{40}
\bibitem [{Not()}]{Note}%
  \BibitemOpen
\bibfield  {journal} {  }\href@noop {} {\bibinfo  {journal} {Ti$^{3+}$ in bulk
  STO can be due to residual defects and/or long living photodoped carriers
  promoted from shallow energy traps under the beam irradiation}\ }\BibitemShut
  {NoStop}%
\end{thebibliography}
\end{document}